# Energy-Efficient Moderate Precision Time-Domain Mixed-signal Vector-by-Matrix Multiplier Exploiting 1T-1R Arrays

Shubham Sahay, *Member, IEEE,* Mohammad Bavandpour, Mohammad Reza Mahmoodi, and Dmitri Strukov, *Senior Member, IEEE*

*Abstract*— **The emerging mobile devices in this era of internet-of-things (IoT) require a dedicated processor to enable computationally intensive applications such as neuromorphic computing and signal processing. Vector-by-matrix multiplication (VMM) is the most prominent operation in these applications. Therefore, there is a critical need for compact and ultralow-power VMM blocks to perform resource-intensive low-to-moderate precision computations. To this end, in this work, for the first time, we propose a time-domain mixed-signal VMM exploiting a modified configuration of 1MOSFET-1RRAM (1T-1R) array. The proposed VMM overcomes the energy inefficiency of the current-mode VMM approaches based on RRAMs. A rigorous analysis of the different non-ideal factors affecting the computational precision indicates that the non-negligible minimum cell currents, channel length modulation (CLM) and drain-induced barrier lowering (DIBL) are the dominant mechanisms degrading the precision of the proposed VMM. Our results also indicate that there exists a trade-off between the computational precision, dynamic range, and the area- and energy-efficiency of the proposed VMM approach. Therefore, we provide the necessary design guidelines for optimizing the performance. Our preliminary results show that an effective computational precision of 6–bits is achievable owing to an inherent compensation effect in the modified 1T-1R blocks. Furthermore, a 4-bit 200×200 VMM utilizing the proposed approach exhibits a significantly high energy efficiency of ~1.5 POps/J and a throughput of 2.5 TOps/s including the contribution from the input/output (I/O) circuitry.**

## I. INTRODUCTION

The traditional digital processors are extremely energy inefficient while handling high-dimensional data from operations such as object/speech recognition, image processing, probabilistic inference, etc. [1]-[2]. Moreover, the widespread use of computationally intensive applications such as deep neural networks(DNNs)/recurrent neural networks (RNNs), real-time signal processing and optimization algorithms in this era of internet-of-things (IoT) necessitates the development of dedicated processing blocks within the mobile devices. The vector-by-matrix multipliers (VMMs) form the most integral part (and often the bottleneck) of these computationally intensive systems. Therefore, the development of a compact and energy-efficient VMM engine is highly essential [3]-[16].

The analog-domain VMM implementations are more area- and energy-efficient as compared to the digital counterparts for computational tasks such as inference, classification, recognition, etc. which are robust to low resolution (reduced precision) VMM operations (and can be trained effectively to handle hardware imperfections without compromising with the accuracy) [3], [6]-[10]. Recently, VMMs based on emerging non-volatile memories, RRAMs in particular, have attracted considerable attention since the VMM operation is simplified as current accumulation through programmable resistances in analog domain [5]-[6], [10]. However, the current-mode VMM implementations based on RRAM require high current levels [6], [16] and bulky trans-impedance amplifiers at each column of the cross-bar [6], resulting in a significant area- and energy-overhead. Moreover, the computational precision is also limited in such implementations and may only be improved at the cost of an increased area to accommodate complex peripheral circuitry implementing sophisticated tuning algorithms or non-trivial mapping techniques [6].

Recently, energy-efficient time-domain VMMs [4], [9]-[15] exploiting post-synaptic pulse (PSP) emulators [11], SRAM (binary) outputs [13], 2D-NOR flash [15] and 3D-NAND flash memory [20] as programmable weights have also been proposed. The energy efficiency of even RRAM-based VMM approaches could be significantly improved if such a time-domain switched-capacitor based approach [8] is followed as opposed to the power-hungry current-mode approach. Furthermore, while the high drain-induced barrier lowering (DIBL) in the 2D-NOR flash memory owing to the poor gate control due to a higher effective oxide thickness (EOT) leads to an increased computational error [15], the significant capacitive-coupling between the bit-line (BL) and the bit-select transistor (BSL) in the 3D-NAND flash memory and the consequent charge-disturbance error restricts their computational precision [20]. The utilization of 1Transistor-1RRAM (1T-1R) cells consisting of MOSFETs, which exhibit an enhanced electrostatic integrity due to a lower EOT and

---

The authors are with the California Nano Systems Institute (CNSI) and also with the Department of Electrical and Computer Engineering, University of California, Santa Barbara, California, 93106, U.S.A.

(e-mail: shubhamsahay@ucsb.edu; strukov@ece.ucsb.edu)

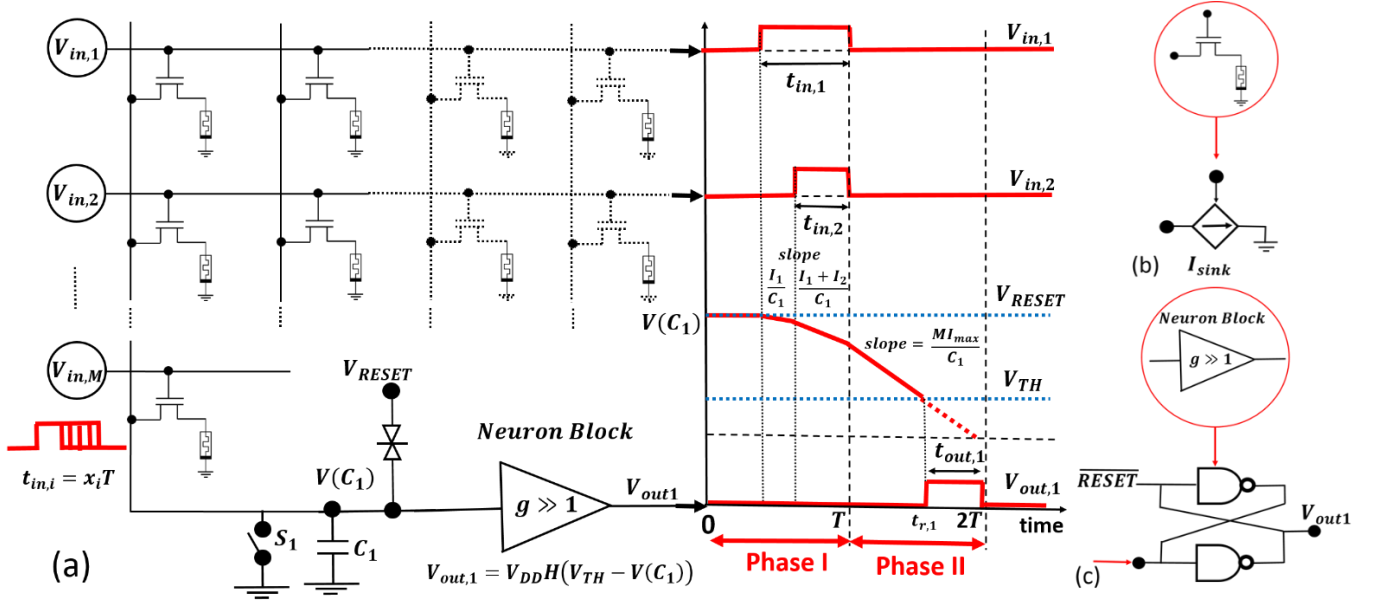

Fig.1 Schematic view of (a) the VMM circuit utilizing 1T-1R array and the timing diagram of the inputs, outputs and the voltage across the load capacitor, (b) the modified 1T-1R block which acts as a programmable current sink and (c) the peripheral circuit within the neuron block implementing the Heaviside function.

significantly reduced capacitive-coupling due to a low gate-drain capacitance, may overcome these limitations and facilitate realization of VMMs with higher computational precision. To this end, in this work, for the first time, we propose a time-domain mixed-signal VMM exploiting a modified 1MOSFET-1RRAM (1T-1R) array. In the proposed VMM approach, the weights are realized as programmable current sinks via tuning the conductance state of the RRAM in the modified 1T-1R blocks in the analog domain while the inputs and outputs are encoded as pulse durations in the digital domain. Contrary to the conventional 1T-1R blocks, where RRAM is connected to the drain of the MOSFET, the RRAM is attached to the source in this approach. This leads to an inherent negative feedback which we call the "self-compensation effect" and significantly improves the computational precision. We also performed a rigorous analysis of the different factors such as drain-induced barrier lowering (DIBL)/channel length modulation (CLM), capacitive coupling, non-negligible minimum cell currents, etc. which may degrade the computational precision. Our preliminary results show that an effective computational precision of 6-bits and an energy efficiency of ~1.5 Pops/J and a throughput of 2.5 Tops/s for a 4-bit 200×200 VMM may be achieved utilizing this approach.

The paper is organized as follows: the proposed VMM approach is discussed in section II. The impact of RRAM behavior on the characteristics of the modified 1T-1R cells and the different physical mechanisms which may affect the performance of the proposed approach are discussed in section III. The design guidelines for optimizing the performance of the proposed 1T-1R VMM are discussed in section IV and the area, energy and throughput estimates are provided in section V. The detrimental effect of the non-negligible minimum cell currents and the design methodology to tailor the proposed VMM for lower output currents are discussed in section VI and the conclusions are drawn in section VII.

## II. PROPOSED VMM APPROACH

A generalized $M \times N$ VMM operation may be represented as:
$$y_j = \frac{1}{M}\sum_{i=1}^{M} w_{ij}x_i, \quad j = 1,2,\ldots,N \qquad (i)$$
where the inputs $x_i$, outputs $y_j$ and weights $w_{ij}$ are normalized such that $(x_i, y_j, w_{ij}) \in [0,1]$. The proposed time-domain VMM approach exploiting the modified 1T-1R array is shown in Fig. 1. In the time-domain VMM [9]-[], the inputs are encoded as duration of the digital pulses such that:
$$t_{in,i} = x_i T \qquad (ii)$$
where $T$ is the time window for the VMM operation. In the proposed approach, the modified 1T-1R block acts as a programmable current sink as shown in Fig. 1(b) and the digital inputs applied to the gate of the MOSFETs ($V_{in,i}$) enable the $i^{th}$ current sink for a duration $t_{in,i}$. It may be noted that unlike conventional 1T-1R arrays where the RRAMs are connected to the drain of the MOSFETs, in this approach, the RRAMs are connected to the source of the MOSFETs. This modification dissuades the non-idealities such as channel length modulation (CLM) and drain induced barrier lowering (DIBL) owing to the self-compensation effect as discussed in section III.C. The weights ($w_{ij} \in [0,1]$) are mapped to the currents ($I_{ij} \in [I_{min}, I_{max}]$) through the programmable current sink as:
$$I_{ij} = I_{min} + w_{ij}(I_{max} - I_{min}) \qquad (iii)$$
Each column of the programmable current sinks is connected to a load capacitor $C_j$. A threshold (neuron) circuit proposed in [14] with a transfer function given as:
$$V_{outj} = V_{DD}H\left(V_{TH} - V(C_j)\right) \qquad (iv)$$
where $H()$ is the Heaviside function encodes the voltage on load capacitor $C_j$ into output digital pulse duration.

The entire VMM operation is completed in two cycles (phase-I and phase-II) of duration $T$ each. The load capacitor $C_j$ is initially pre-charged to a voltage $V_{RESET}$ at the beginning of phase-I ($t = 0$). The inputs are activated only in phase-I (integration phase) and the current sinks start discharging $C_j$. At the end of phase-I ($t = T$), the voltage across the load capacitor $V(C_j)$ reduces by $\Delta V(C_j)$ where:

$$\Delta V(C_j)_{t=T} = \frac{1}{C_j}\sum_{i=1}^{M} I_{ij} t_{in,i} \quad \text{(v)}$$

Using the expression for $I_{ij}$ from equation (iii) in equation (v), we get:

$$\Delta V(C_j)_{t=T} = \frac{T(I_{max}-I_{min})}{C_j}\sum_{i=1}^{M} w_{ij} x_{in,i} + \frac{TI_{min}}{C_j}\sum_{i=1}^{M} x_{in,i} \quad \text{(vi)}$$

As evident from equation (vi), the change in the voltage across the load capacitor at the end of phase I is mapped to a linear expression of the weighted sum in this scheme. To ensure that this voltage variation across the load capacitor is limited to a targeted operation regime, i.e. $V(C_j)_{t=T} \in [V_{RESET}, V_{TH}]$, the load capacitor $C_j$ must be designed such that:

$$C_j = \frac{MI_{max}T}{V_{RESET}-V_{TH}} \quad \text{(vii)}$$

In phase-II (evaluation phase), all the inputs are inactivated and the load capacitor is discharged through a constant current $MI_{max}$. This discharging current may be generated either via a current mirror or by adding a similar 1T-1R array at the load capacitor with all the inputs activated for the entire duration $T$ during phase-II and all the current sinks programmed to $I_{max}$. In this work, we have followed the latter approach to implement the constant current source during phase-II. The neuron circuit generates an output pulse when the voltage on the load capacitor reaches the threshold voltage i.e. ($V(C_j) = V_{TH}$). The time instance ($t_{r,j}$) at which $V(C_j) = V_{TH}$ can be given as:

$$t_{r,j} = T - t_{out,j} = T\left[1 - \frac{\sum_{i=1}^{M} I_{ij} t_{in,i}}{MI_{max}T}\right] \quad \text{(viii)}$$

The output pulse duration ($t_{out,j}$) can be simply obtained by using equations (i) and (iii) in equation (viii) as:

$$t_{out,j} = ay_j T + b \quad \text{(ix)}$$

where,

$$a = \frac{(I_{max}-I_{min})}{I_{max}}, \quad b = \frac{I_{min}}{MI_{max}}\sum_{i=1}^{M} x_{in,i}. \quad \text{(x)}$$

Equation (ix) clearly indicates that the output result obtained using the proposed scheme is different from the targeted ideal output result ($t_{out,j} = y_j T$) due to the non-zero (appreciable) minimum current ($I_{min}$) of the 1T-1R cells which lead to the undesirable multiplicative coefficient ($a$) and the input-dependent additive coefficient ($b$).

However, it may be noted that the input-dependent additive coefficient ($b$) can be cancelled out by utilizing the differential scheme. In the differential implementation, each weight is realized utilizing two sub-weights $w_{ij}^+$ and $w_{ij}^-$ such that

$$w_{ij} = w_{ij}^+ - w_{ij}^- \quad \text{(xi)}$$

and two sub-neurons are dedicated to calculate the dot product of inputs and each sub-weight vector as $t_{out,j}^+$ and $t_{out,j}^-$ as:

$$t_{out,j}^+ = \frac{a}{M}\sum_{i=1}^{M} w_{ij}^+ x_i + b \quad \text{(xii)}$$

$$t_{out,j}^- = \frac{a}{M}\sum_{i=1}^{M} w_{ij}^- x_i + b \quad \text{(xiii)}$$

A simple logic circuitry is then employed to generate the final differential output pulse as:

$$t_{out,j} = t_{out,j}^+ - t_{out,j}^- \quad \text{(xiv)}$$

On the other hand, the multiplicative coefficient ($a$) leads to a reduction in the output time window. This shrinkage can be compensated by either lowering the constant current during the evaluation phase (which extends the time window for phase II), or increasing the output time-to-digital convertor (TDC) counter frequency. Furthermore, the non-negligible minimum cell currents also lead to a reduction in the portion of the output swing available for performing useful computations as discussed in section VI.

## II. 1T-1R VMM DESIGN GUIDELINES

The performance of the proposed 1T-1R VMM was evaluated at the 55-nm technology node using process design kit (PDK) from Global Foundries in HSPICE (version N-2017.12[17]). Furthermore, a rather simplistic compact model was used for RRAM with the current-voltage relationship expressed as:

$$I_{mem} = \frac{1}{\beta R_0}\sinh(\beta V_{mem}) \quad \text{(xv)}$$

where $R_0$ is the low-voltage resistance, and $\beta$ is the non-linearity factor [18]. A maximum ON-state resistance ($R_{ON}$) of 2.5 KΩ and a minimum OFF-state resistance ($R_{OFF}$) of 2.5 MΩ were considered for RRAM to explore the entire design space. Furthermore, a maximum permissible read voltage of 0.5 V without disturbing the programmed state of RRAM was assumed. Under these assumptions, we evaluated the potential of the proposed 1T-1R time-domain VMM under different operating conditions and different parameters for the RRAM. In the subsequent sections, we discuss the operating conditions and provide the necessary design guidelines to extract the optimum performance from the proposed VMM architecture. It may be noted that the optimal conditions also differ with the input constraints such as VMM size, input voltage, time window, dynamic range ($DR = I_{max} - I_{min}$), targeted precision, etc.

### A. Impact of RRAM behavior on 1T-1R cell characteristics

The load-line characteristics of the 1T-1R block (MOSFET with minimum gate length, $L_g$ = 60 nm and minimum width, $W$ = 120 nm) is shown in Fig. 2 for different non-linearity factors of the RRAM (β). The reset voltage $V_{RESET}$ was chosen as 0.9 V to reduce the error induced due to non-idealities such as CLM and DIBL (as discussed in section III.C). As shown in Fig. 2, for a particular gate voltage, the range of drain voltages at which the 1T-1R cells may operate is limited between the intersection of the RRAM characteristics with extreme resistance values ($R_{ON}$ and $R_{OFF}$) and the output characteristics of the MOSFET. We observe from Fig. 2 that increasing the non-linearity factor of the RRAM results in a reduction of the operating regime for low gate voltages (< 0.6 V) i.e. near-threshold and sub-threshold operating regimes. A reduced operating range of drain voltage leads to a lower *DR* which may be obtained from the modified 1T-1R block via tuning the conductance state of the RRAM (absolute values are reported in Table I).

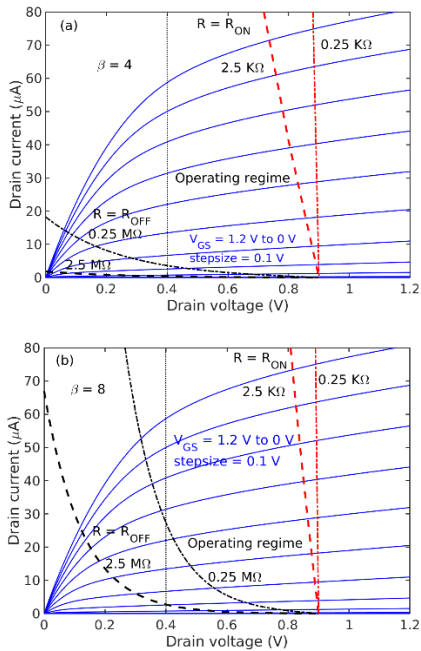

Fig. 2 The load line characteristics of the modified 1T-1R block for RRAMs with different non-linearity factor (a) β = 4 and (b) β = 8.

Moreover, the operating range of drain voltages is also degraded when a RRAM with a higher $R_{ON}$ or a lower $R_{OFF}$ is used as shown in Fig. 2. Also, the ON-state to OFF-state conductance ratio of the RRAM should be high to obtain an appreciable $DR$. Furthermore, unlike the current-mode VMM approach based on RRAM, where the accumulated current depends exclusively on the conductance state of the RRAMs, the cell currents of the modified 1T-1R block depends both on the conductance state of the RRAM and the channel conductance of the MOSFET (which can also be tuned by the input gate voltage). Therefore, even if we utilize a RRAM with higher ON-state conductance (for instance, 10 times smaller $R_{ON}$), the cell current of the 1T-1R block increases only slightly (< 2 times) as shown in Fig. 2. Therefore, the energy-efficiency (dictated by the cell currents) of the proposed VMM approach is not degraded considerably even when RRAMs with low ON-state resistance are used as opposed to the current-mode VMM where the accumulated current would increase significantly (by a decade for the former case) and limit the energy-efficiency. Moreover, the $DR$ also increases for lower $R_{ON}$ in the 1T-1R configuration. Therefore, the proposed modified 1T-1R blocks may facilitate energy-efficient VMM operations utilizing RRAMs with low ON-state resistances and high $R_{OFF}/R_{ON}$ ratio.

*B. Precision*

The effective weight precision (i.e. programmability of the current sinks) depends on the accuracy of tuning the conductance states of RRAM and degrades due to a drift in the analog conductance state with cycling, temperature and the inherent intrinsic noise such as RTN. Previous works have already shown an effective weight precision greater than 7-bits based on a simple tuning algorithm [19]. The weight precision may be further improved by oxide material engineering or by utilizing more efficient tuning algorithms.

As discussed in [15], the computational error (or output error, $e_{outj}$) may be decoupled from the weight error and defined separately as the maximum difference between the theoretically calculated output time period considering ideal current sinks ($t_{outj}^{cal}$ from equation (viii)) and the output time period obtained via transient simulation of the entire VMM circuit ($t_{outj}^{sim}$) in HSPICE, spanning over the entire sample space of the weights and inputs i.e.

$$e_{outj} = \max_{toutj} \frac{\left|t_{outj}^{cal} - t_{outj}^{sim}\right|}{T} \qquad (xvi)$$

For benchmarking against the digital VMM implementations, the effective computational precision ($P_{outj}$) can then be defined as:

$$P_{outj} = -\log_2 e_{outj} - 1 \qquad (xvii)$$

Considering the efficacy of the differential scheme in cancelling the impact of the input-dependent additive coefficient ($b$) as discussed in section II, improving the noise immunity and enhancing the output precision while enabling inclusion of bipolar weights [8], two adjacent columns of the modified 1T-1R array were tuned for implementing the positive and negative weight components of the bipolar weight matrix. Furthermore, the adjacent neuron circuits were used to calculate the positive ($V_{outj} = t_{out,j}^+$) and negative ($V_{out(j+1)} = t_{out,j}^-$) component of output in this differential implementation. The final output was then obtained as the time difference between the rising edge of the neuron circuits used for obtaining the positive ($V_{outj}$) and negative ($V_{out(j+1)}$) component of the output. This rectified linear (ReLU) operation was implemented utilizing a digital gate for $V_{finalj} = V_{outj} \cdot \overline{V_{out(j+1)}}$.

*C. Non-ideal factors*

The computational precision is degraded by several factors which tend to deviate the current sinks from draining a constant current. While CLM leads to a linear dependence of the MOSFET's drain current on the drain voltage, the DIBL effect induces threshold voltage shift which further increases the variation in the drain current with the drain voltage. Therefore, in addition to the input gate voltage ($V_{GS}$), the current through the programmable 1T-1R current sink also depends on the drain voltage i.e. the output voltage at the load capacitor.

To minimize this dependency of the cell currents on the output voltage, we modified the conventional 1T-1R array architecture. While the RRAM is connected to the drain terminal of the MOSFET in the conventional 1T-1R array, one terminal of RRAM is connected to the source of the MOSFET and the other terminal is grounded in this implementation as shown in Fig. 1(b). An increase in the drain voltage in the modified 1T-1R configuration with RRAM connected to the source leads to an enhanced current flowing through the RRAM. This results in a larger voltage drop across the RRAM. The increased voltage drop across the RRAM provides a negative feedback and effectively boosts the source potential leading to a reduction in the effective gate-to-source voltage ($V_{GS}$) which in turn suppresses the increment in the drain current. Therefore, an increase in the drain current due to the application of a larger drain voltage is compensated by a reduction in the effective gate overdrive voltage in the modified 1T-1R array. This inherent self-compensation effect leads to a diminished dependency of the modified 1T-1R cell currents on the output voltage at the load capacitor.

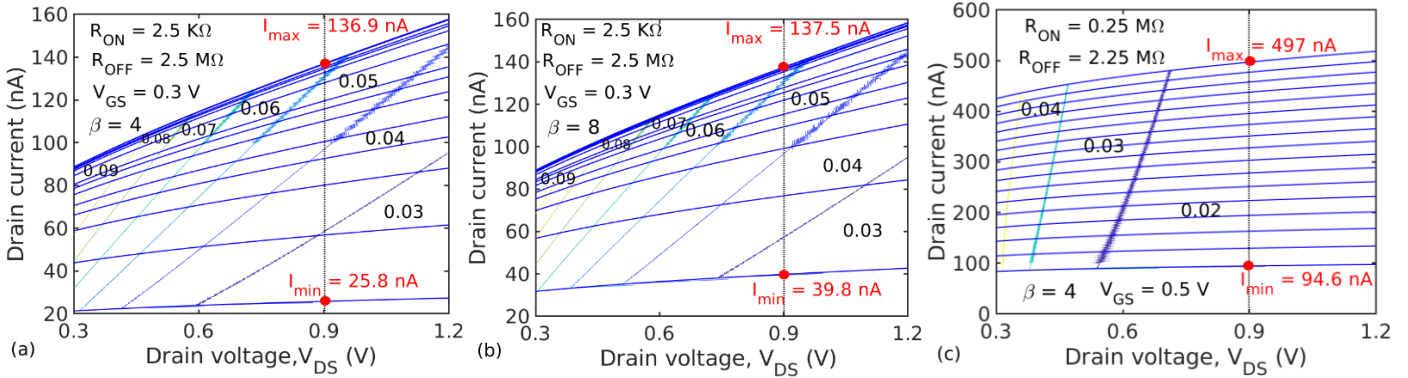

Fig. 3 The error contour plot due to the DIBL and CLM effect for different input voltages and non-ideality factors: (a) $V_{GS} = 0.3$, $\beta = 4$ and (b) $V_{GS} = 0.3$, $\beta = 8$ and (c) $V_{GS} = 0.5$ and $\beta = 4$ for MOSFETs with $L_g = 120$ nm.

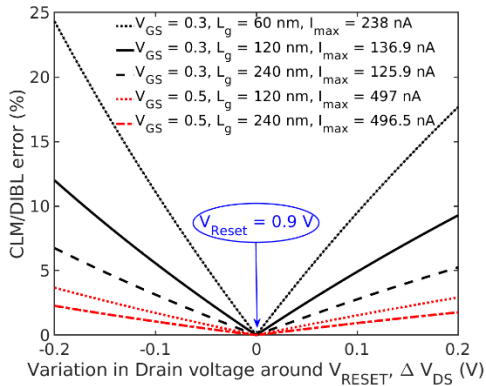

Fig. 4 The total error due to the DIBL and CLM effect for different input voltages and gate lengths ($L_g$) of the MOSFET with non-ideality factor, $\beta = 4$.

The error due to CLM and DIBL can be defined as:

$$e_{CLM/DIBL} = 1 - \frac{I(V - \Delta V)}{I(V)} \qquad (xviii)$$

where $\Delta V$ is chosen as 1 mV to estimate the local error contours with high accuracy (Fig. 3). As can be observed from the error contour plots in Fig. 3, we performed a rigorous analysis of the CLM and DIBL error for different input gate voltages and non-ideality factor ($\beta$) of RRAM within the operating regime of the modified 1T-1R configuration. For all the input gate voltages, we found that the cell currents are relatively independent of the drain voltage (i.e. the CLM/DIBL error is low) for higher drain voltages. Therefore, we selected a high reset voltage, $V_{RESET} = 0.9$ V and designed the neuron circuit to have a threshold voltage $V_{TH} = 0.7$ V to ensure a non-disturbing maximum voltage swing of 0.2 V across the RRAM. Furthermore, we also observe from Fig. 3 that the DIBL/CLM error increases as we reduce the input voltage and operate with a smaller maximum current ($I_{max}$) to limit the load capacitance (see equation (vii)). Also, as discussed in section III.A, an increase in the non-linearity factor of the RRAM leads to a reduction in the dynamic range as shown in Fig. 3(b).

Since the DIBL and CLM mechanisms are extremely sensitive to the gate length ($L_g$) of the MOSFETs, we also performed a thorough investigation of the CLM/DIBL error for 1T-1R cells consisting of MOSFETs with different gate lengths biased at different input voltages as shown in Fig. 4. We found that despite the self-compensation effect, the CLM/DIBL error is significantly high in the MOSFETs with minimum gate length ($L_g = 60$ nm) and reduces drastically by ~5 times when the gate length is quadrupled to $L_g = 240$ nm for the same input voltage. Moreover, the CLM/DIBL error can be further reduced while obtaining a higher $DR$ by increasing the input voltage $V_{GS}$ to 0.5 V at the cost of an increased capacitor area and energy owing to larger $I_{max}$ (Fig. 4). Therefore, there exists a trade-off between area- and energy-efficiency, the dynamic range and the computational error in the proposed approach.

It may also be noted that the DIBL error is lower in the 1T-1R cells as compared to the 2D-NOR flash memory cells. This is attributed to the smaller effective oxide thickness (EOT) and the consequent enhanced electrostatic integrity in the MOSFETs as compared to 2D-NOR flash cells. Therefore, the computational precision is higher in the proposed approach as compared to 2D-NOR flash based time-domain VMM implementation [15].

Apart from the error induced due to CLM and DIBL, the capacitive coupling between the load capacitor and the gate-drain capacitance of the MOSFET could be another possible source of charge disturbance. However, in the proposed architecture, the load capacitor is large as compared to the gate-drain capacitance of the MOSFETs which diminishes the charge disturbance error due to capacitive-coupling. Therefore, the proposed VMM also exhibits a higher precision as compared to the 3D-NAND flash memory based time-domain VMM where the significant coupling between the bit-line (BL) and the bit-select transistor (BSL) leads to a considerable charge disturbance error [20].

Furthermore, the intrinsic thermal noise of the MOSFET and the random telegraph noise (RTN) in the RRAM may also affect the computational precision. Therefore, performance analysis of the proposed VMM under different RRAM noise sources is an important future work.

TABLE I
DESIGN SPACE EXPLORATION

| $V_{GS}$ (V) | 0.3 ($L_g$ = 120 nm, $R_{ON}$ = 2.5 KΩ, $R_{OFF}$=2.5MΩ) | | | | | | 0.3 ($L_g$ = 240 nm, $R_{ON}$ = 2.5 KΩ, $R_{OFF}$=10MΩ) | | | | | | 0.5 ($R_{ON}$ = 0.25 MΩ) | | | | | |
|---|---|---|---|---|---|---|---|---|---|---|---|---|---|---|---|---|---|---|
| B | 4 | | | 8 | | | 4 | | | 8 | | | 4 ($L_g$ = 120 nm, $R_{OFF}$ = 2.25 MΩ) | | | 4 ($L_g$ = 240 nm, $R_{OFF}$ = 2.375 MΩ) | | |
| $I_{max}; I_{min}$ | 136.9 nA ; 25.8 nA | | | 137.5 nA ; 39.8 nA | | | 125.9 nA ; 25.2 nA | | | 126.3 nA ; 38.7 nA | | | 497 nA ; 94.6 nA | | | 496.5 nA ; 94.1 nA | | |
| T (ns) | 16 | 32 | 64 | 16 | 32 | 64 | 16 | 32 | 64 | 16 | 32 | 64 | 16 | 32 | 64 | 16 | 32 | 64 |
| VMM size, M = 10 | | | | | | | | | | | | | | | | | | |
| $E_{Cl}$(pJ) | 0.09 | 0.19 | 0.39 | 0.09 | 0.19 | 0.39 | 0.09 | 0.18 | 0.36 | 0.09 | 0.18 | 0.36 | 0.3 | 0.7 | 1.4 | 0.3 | 0.7 | 1.4 |
| $e_{out}$,% | 4.5 | 2.8 | 2.8 | 4.4 | 2.8 | 2.8 | 2.6 | 2.5 | 2.4 | 2.6 | 2.5 | 2.4 | 1.44 | 1.0 | 0.74 | 0.88 | 0.74 | 0.55 |
| $P_{out}$ | 3 | 4 | 4 | 3 | 4 | 4 | 4 | 4 | 4 | 4 | 4 | 4 | 5 | 5 | 6 | 5 | 6 | 6 |
| VMM size, M = 50 | | | | | | | | | | | | | | | | | | |
| $E_{Cl}$(pJ) | 2.45 | 4.92 | 9.85 | 2.47 | 4.95 | 9.9 | 2.25 | 4.53 | 9.06 | 2.27 | 4.5 | 9.09 | 8.93 | 17.8 | 36 | 8.95 | 17.8 | 36 |
| $e_{out}$,% | 4.3 | 2.6 | 2.7 | 4.2 | 2.7 | 2.6 | 2.4 | 2.4 | 2.3 | 2.5 | 2.4 | 2.2 | 1.3 | 0.94 | 0.72 | 0.77 | 0.68 | 0.48 |
| $P_{out}$ | 3 | 4 | 4 | 3 | 4 | 4 | 4 | 4 | 4 | 4 | 4 | 4 | 5 | 5 | 6 | 5 | 6 | 6 |
| VMM size, M = 100 | | | | | | | | | | | | | | | | | | |
| $E_{Cl}$(pJ) | 9.81 | 19.7 | 39.4 | 9.9 | 19.8 | 39.6 | 9.0 | 18.1 | 36.2 | 9.09 | 18.2 | 36.3 | 35.7 | 71.5 | 144 | 35.6 | 71.4 | 144 |
| $e_{out}$,% | 4.2 | 2.6 | 2.6 | 4.2 | 2.6 | 2.5 | 2.3 | 2.3 | 2.3 | 2.3 | 2.4 | 2.2 | 1.2 | 0.92 | 0.66 | 0.75 | 0.64 | 0.46 |
| $P_{out}$ | 3 | 4 | 4 | 3 | 4 | 4 | 4 | 4 | 4 | 4 | 4 | 4 | 5 | 5 | 6 | 6 | 6 | 6 |
| VMM size, M = 200 | | | | | | | | | | | | | | | | | | |
| $E_{Cl}$(pJ) | 39.2 | 78.4 | 157 | 39.6 | 79.2 | 158 | 36 | 72.5 | 145 | 36.3 | 72.7 | 145 | 142 | 286 | 576 | 142 | 285 | 576 |
| $e_{out}$,% | 4.3 | 2.7 | 2.7 | 4.4 | 2.7 | 2.7 | 2.4 | 2.4 | 2.3 | 2.4 | 2.3 | 2.3 | 1.3 | 0.93 | 0.67 | 0.77 | 0.64 | 0.46 |
| $P_{out}$ | 3 | 4 | 4 | 3 | 4 | 4 | 4 | 4 | 4 | 4 | 4 | 4 | 5 | 5 | 6 | 6 | 6 | 6 |

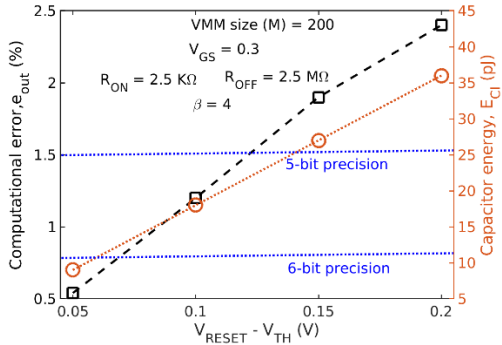

Fig. 5 Impact of variation in the threshold voltage of the neuron circuit ($V_{TH}$) to limit the maximum voltage swing across RRAM ($V_{RESET} - V_{TH}$) on the compute error and the capacitor energy of the proposed VMM approach.

## III. DESIGN SPACE EXPLORATION

We performed a rigorous analysis to explore the design space for optimizing the performance of the proposed VMM architecture. The input gate voltage ($V_{GS}$) and time window ($T$) are the most crucial design parameters for tuning the performance of the proposed VMM for a particular gate length of the MOSFET utilized in the 1T-1R block. The performance-metrics for the proposed VMM with different input voltages ($V_{GS}$), time window ($T$), gate lengths ($L_g$), VMM sizes ($M$ in $M \times M$ VMM) and non-linearity factor ($\beta$) of RRAM are listed in Table I.

The output (worst case) error ($e_{out}$) was found by simulating multiple runs of VMM operation in HSPICE with different combination of random inputs and random weights in each run in an attempt to span the entire sample space of possible input and weight combinations. The line parasitics such as line resistances and capacitances and the corresponding process variations pertinent to the 55-nm technology node were also considered in the HSPICE simulations. The total energy dissipated in the load capacitor, $E_{Cl}$ (which is the dominant energy dissipation mechanism as discussed later in section V) for the VMM operation has also been included in Table I. As can be observed from table I, the output error reduces with increasing VMM size till $M < 100$. However, as the VMM size increases above 100, the line parasitics and their process variations lead to a non-negligible increase in the computational error. While the line resistances lead to a drop in the effective input (gate) voltage of the MOSFETs on the far end of the 1T-1R array leading to a reduced drain current, the line capacitances add to the latency. Although the differential configuration is effective in mitigating the impact of fixed line parasitics, the process variations cannot be compensated even exploiting a differential configuration and escalate the compute error.

From Table I, it can also be observed that there is a trade-off between the computational precision, dynamic range and the energy dissipated in the load capacitor. For instance, to achieve a high computational precision of ~6-bits for large sized VMMs ($M > 100$), a higher value of input voltage ($V_{GS}$ = 0.5 V) should be used. A higher input voltage results in a higher maximum current ($I_{max}$) leading to a larger load capacitance. Although the dynamic range is also high for such operating conditions, the area- and energy-efficiency is limited by the load capacitor which dominates the area and energy landscape (as discussed in Section V.)

Moreover, to achieve a higher area- and energy-efficiency by limiting the size of the load capacitor, a lower input voltage may be used to reduce the maximum current ($I_{max}$). However, the computational precision and the dynamic range reduces significantly at such operating conditions. The weight precision may also limit the computational precision in such cases.

Still, the preliminary results indicate that an effective computational precision of 6-bits is achievable for a VMM size, $M > 100$ using the proposed approach. In addition, depending on the targeted precision, input time window, VMM size, area, energy efficiency, voltage swing across RRAM etc. we may optimize the design parameters to achieve optimum performance of the proposed VMM architecture.

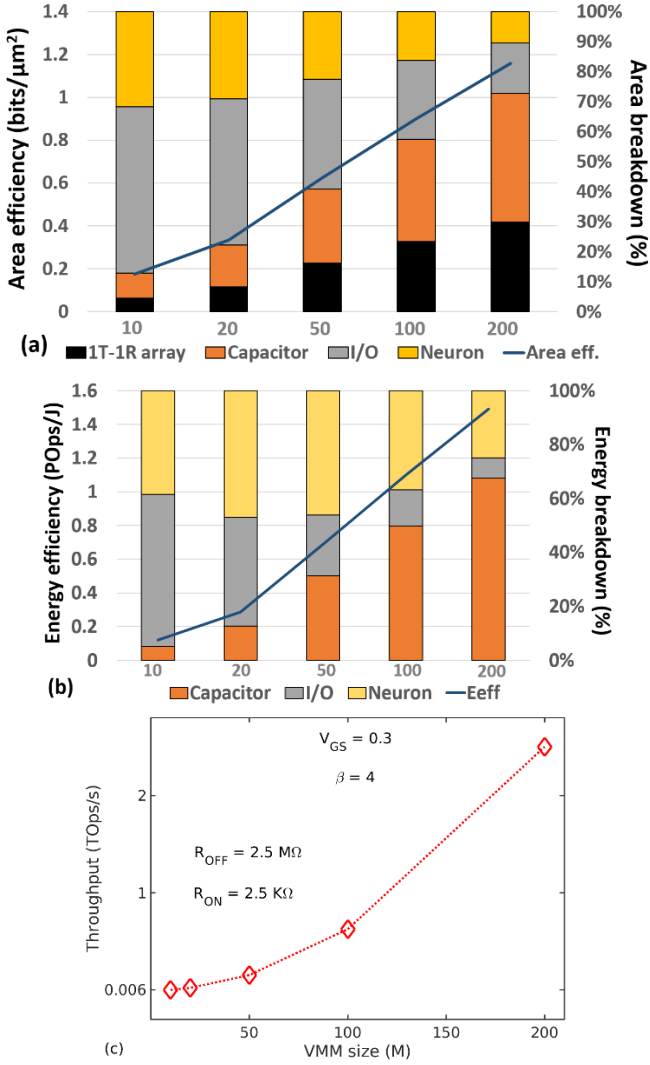

Fig. 6 The variation of (a) area efficiency, (b) energy efficiency and (c) throughput of the proposed VMM with VMM size ($M$) for a ReLU neuron.

Since the conductance-state of the RRAM is sensitive to the voltage drop across them, we have also analyzed the performance of the proposed VMM approach for neuron circuit with different threshold voltages ($V_{TH} > 0.5$ V) to limit the maximum voltage swing across RRAM ($V_{RESET} - V_{TH}$). As can be observed from Fig. 5, a reduction in the maximum voltage swing across RRAM leads to a higher computational precision owing to the lower CLM/DIBL error. Although a reduction in the voltage drop across RRAM increases the load capacitor size according to equation (vii), the energy dissipated in the load capacitor, $E_{Cl}$ decreases owing to the reduced voltage swing as shown in Fig. 5.

## V. PERFORMANCE ESTIMATION

It can be observed from table I that the proposed VMM approach yields a computational precision of 3-bits to 6-bits depending on the design parameters. Targeting a VMM engine with precision of 4-bits, which is sufficient for several applications including neuromorphic computing [8], [10], we select an input voltage $V_{GS} = 0.3$ V, a time window $T = 16$ ns and a gate length $L_g = 240$ nm for estimating the area- and energy-efficiency of the proposed approach. Fig. 6 shows the area and energy breakdown of the proposed VMM considering the input/output (I/O) peripheral circuitry as well as the neuron circuitry for different VMM sizes.

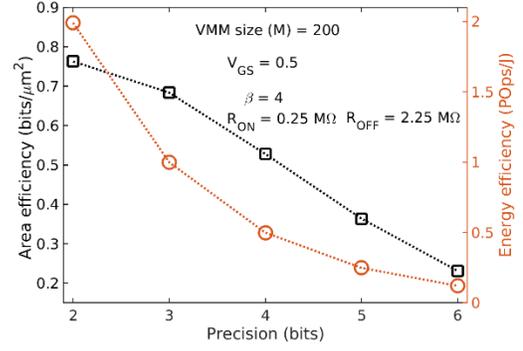

Fig. 7 The variation of area efficiency and energy efficiency of the proposed VMM with VMM size, $M = 200$ for different targeted precisions.

The basic components of the VMM I/O circuitry are digital input to time-domain pulse converters (DTC) which consist of a 4-bit shared counter and a 4-bit digital comparator followed by a S-R latch for each input, and time-domain pulse to digital output converters (TDC) which consist of a 4-bit accumulator for each neuron output [20]. The 4-bit accumulator is realized using a 4-bit full adder and a 4-bit register based on D-flip flops. A shared clock enables conversion of the pulse duration of the neuron output to digital outputs. The neuron circuit consists of a S-R latch realized using a pair of NAND gates followed by an AND gate and NOT gate for implementing the differential scheme [20]. The load capacitors are realized using MOSCAPs from the 55-nm technology node.

It can be observed from Fig. 6 that the I/O circuitry consumes a significant portion of the energy and area landscape of the proposed VMM architecture when the VMM size is small. However, the load capacitor ($C_j$) tends to dominate the area and energy landscape as the VMM size increases. The preliminary results indicate an energy-efficiency of ~1.5 Pops/J and a throughput of 2.5 Tops/s for a 4-bit 200×200 VMM engine utilizing the proposed approach.

Although applications such as inference, classification, recognition etc. may be performed utilizing even low precision (~4 bits) VMM engines [8] without significantly degrading the accuracy, we also analyze the efficacy of the proposed approach for binary-to-moderate (2-bits to 6-bits) target precisions under a different operating condition ($V_{GS} = 0.5$ V) as shown in Fig. 7. An increase in the targeted precision effectively translates into a larger time window ($T$) to encode the inputs while operating at the same frequency. Therefore, the capacitor and I/O circuit area and energy consumption increases significantly with an increased target precision. This leads to a considerable degradation in the area and energy efficiency when targeting moderate precision (> 4-bits) VMM operations as shown in Fig. 7. However, the proposed VMM still exhibits a significantly high energy-efficiency of ~123.1 TOps/J and a throughput of 0.63 Tops/s for 6-bit 200×200 VMM operation.

Moreover, we may utilize RRAMs with a higher ON-state resistance to reduce $I_{max}$ further and decrease the load capacitance $C_j$ for enhancing the area- and energy-efficiency. Similarly, a lower reset voltage ($V_{RESET}$) may further enhance the VMM performance metrics. The capacitor area may also be reduced by using a different input encoding scheme whereby the individual input bits are encoded as discrete binary pulses and employing successive integration and re-scaling technique to reduce the charge integrated on the load capacitor at the cost of reduced computational precision [21].

## VI. DESIGN MODIFICATION FOR LOWER OUTPUT CURRENTS

In the proposed VMM design, for a VMM size = $M$, the maximum current which is integrated at the load capacitor is $M \times I_{max}$, when all the inputs and weights are maximum (all current sinks are programmed to $I_{max}$). Moreover, from a digital circuit perspective, such a scenario is equivalent to rounding the full precision i.e. $2P + \log_2 M$ bit-long VMM output (obtained by multiplying $M$-numbers with $P$-bit precision in the digital domain) to the most significant $P$-bits where $P$ is the precision of the proposed VMM. However, in some neural networks and several other applications, all the inputs and weights do not attain their maximum value during the operation and the maximum VMM output current is significantly lower than $M \times I_{max}$. Therefore, the VMM design may be further modified according to the expected maximum dot-product value (which translates to a maximum allowable output current in the proposed VMM). Such a modification would not only lead to a reduction in the rounding/quantization error but also facilitate utilization of a smaller load capacitor to integrate the relatively lower output current.

However, a reduction in the maximum output current and the load capacitor leads to an increase in the computational error and significantly degrades the precision of the VMM as shown in Fig. 8(a). The computational precision degrades to 4-bits when the maximum VMM output current is limited to $M^{\frac{1}{2}} \times I_{max}$ (which is equivalent to extracting $P$-bit VMM output from center of the $2P + \log_2 M$ bit-long digital output) and to 2-bits when the maximum VMM output current is reduced to $M^{\frac{1}{3}} \times I_{max}$.

Apart from introducing a multiplicative coefficient ($a$) and the input-dependent additive coefficient ($b$) in the output, the appreciable minimum current of the 1T-1R cells also consumes a significant portion of the output voltage swing. Therefore, the part of the total output swing available for performing useful computation is also low for the proposed VMM due to the non-negligible $I_{min}$. Moreover, a reduction in the maximum output current simply implies that the part of the output voltage swing dominated by the minimum cell currents would be even larger. Furthermore, a reduction in the load capacitor may lead to a charge disturbance error [20] owing to the increased coupling (and hence, charge sharing) between the gate-drain capacitance of the MOSFET and the load capacitor. As a result, the useful portion of the available output swing is further degraded by the charge disturbance error. This leads to a significant reduction in the VMM output precision observed in Fig. 8(a).

Although increasing the output voltage swing may appear as a straightforward technique to reduce the output error and

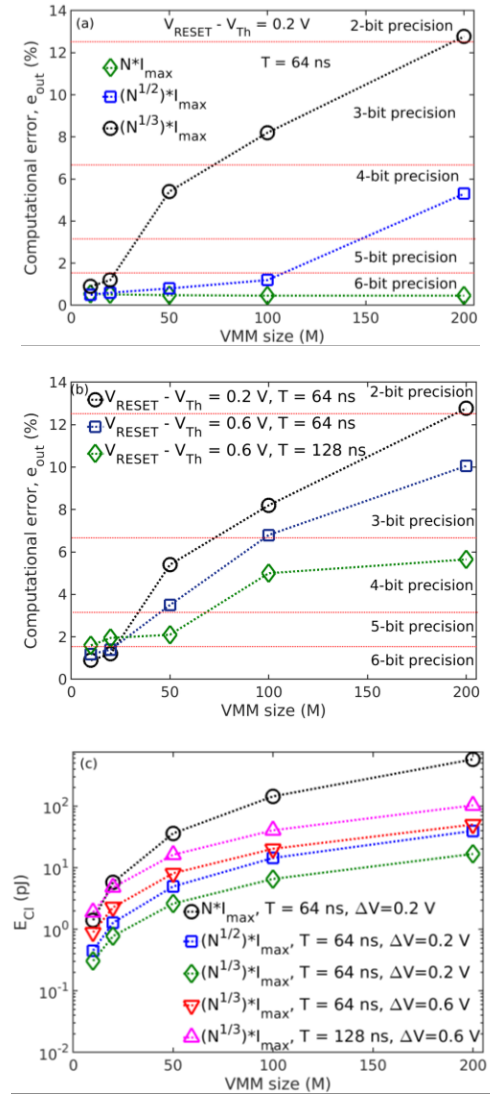

Fig. 8 The computational error of the proposed VMM for different maximum output currents with (a) fixed output swing ($V_{RESET} - V_{TH} = 0.2$ V), (b) the computational error when the maximum current is $M^{\frac{1}{3}} \times I_{max}$ for different output swings ($\Delta V = V_{RESET} - V_{TH}$) and time windows ($T$) and (c) the dominant energy dissipated in the load capacitor for all the cases presented in parts (a) and (b).

improve the precision, it may lead to a degradation in the computational precision when the DIBL/CLM error is high as shown in Fig. 5 (for input voltage $V_{GS} = 0.3$ V). However, when an input voltage, $V_{GS} = 0.5$ V is utilized, DIBL/CLM error is minimized (Fig. 4). In this case, an increase in the output swing enhances the useful portion available for computation leading to an improved computational precision. Moreover, an increase in the output voltage swing further minimizes the load capacitor (refer equation (vii)). This leads to a reduction in the energy dissipated in the load capacitor despite an increase in the voltage swing as shown in Fig. 8(c). However, the computational precision is only 3-bits for the maximum output current of $M^{\frac{1}{3}} \times I_{max}$, even when the output swing is increased to the maximum permissible read voltage, which is not useful for applications such as DNNs. To increase the computational precision further (to 4-bits which is appropriate for applications

TABLE II
PERFORMANCE BENCHMARKING

| Reference | [3] | [6] | [7] | [8] | [10] | [11] | [14] | [20] | [21] | This work |
|---|---|---|---|---|---|---|---|---|---|---|
| **Approach** | CM | CM | CM | SC | TD | TD | TD | TD | TD | TD |
| **Process(nm)** | 180 | 22 | 180 | 40 | 14 | 250 | 55 | 55 | 55 | 55 |
| **Precision (bits)** | 3 | ~4 | ~5 | 3 | <8 | ~1/5* | ~6 | ~5 | ~4 | 6 |
| **EE(Tops/J)** | 6.4 | 60 | 5.7 | 8 | 18 | <290 | 85 | 91 | 1305 | 1496 |
| **I/O included** | Yes | No | Yes | Yes | No | No | Yes | Yes | Yes | Yes |
| **Results** | Sim | Sim | Exp | Exp | Sim | Sim | Sim | Sim | Sim | Sim |

CM: current-mode   SC: switch-capacitor   TD: time-domain   *Binary weights/analog output such as DNNs), a larger time window ($T$) should be utilized as shown in Fig. 8(b). However, a larger time window not only reduces the operating frequency but also increases the size of the load capacitor (refer equation (vii)) and its energy dissipation (Fig. 8(c)). Therefore, there is an inherent trade-off between the computational precision and the energy dissipated in the load capacitor even when the maximum output current is limited to $M^{\frac{1}{3}} \times I_{max}$.

## VII. CONCLUSION

An energy-efficient time-domain VMM exploiting a modified configuration of 1T-1R array has been proposed in this work. The dominant mechanisms such as CLM, DIBL, etc. which degrade the performance of the proposed architecture are discussed in detail. Furthermore, we show that there exists a trade-off between the computational precision, dynamic range and the area- and energy-efficiency of the proposed VMM approach. Therefore, we also provide necessary design guidelines to further optimize the performance of the 1T-1R VMM. The preliminary results indicate that an effective computational precision of 6–bits and a significantly high energy efficiency of ~1.5 POps/J and a throughput of 2.5 Tops/s as compared to the other VMM approaches (Table II) is achievable for a 4-bit 200×200 VMM using the proposed approach. Our results may provide an incentive for experimental realization of the VMM approach based on 1T-1R array.